\documentclass{book} \usepackage{amsmath} \usepackage{hhline}
\usepackage{graphicx} \usepackage[T2A]{fontenc}
\usepackage[cp1251]{inputenc} \usepackage[russian]{babel}

\begin{document}

\def\ds{\displaystyle} \def\ss{\scriptsize} \def\hh{\hskip 1pt}
\def\hs{\hskip 2pt} \def\h{\hskip 0.2mm} \def\pr{\prime}

\newcommand{\fbR}{\hbox{\ss\bf R}}
\newcommand{\fbr}{\hbox{\footnotesize\bf r}}
\newcommand{\fbk}{\hbox{\footnotesize\bf k}}
\newcommand{\Tr}{\mathop{\rm T\h r}\nolimits}

\parindent=5mm

\setcounter{page}{1}

\phantom{X}

\makeatletter\renewcommand{\@evenhead} {\vbox{\hbox
to\textwidth{\rm\thepage\hfil\it Новая теория сверхтекучести \hfil}\vskip 1mm\hrule
\vskip -3mm}}

\renewcommand{\@evenhead}{}

\makeatletter\renewcommand{\@oddhead} {\vbox{\hbox
to\textwidth{\hfill\it Бондарев Б.В.\hfill\rm\thepage}\vskip 1mm\hrule\vskip -3mm}}

\renewcommand{\@oddhead}{}

\vskip 30mm

\centerline {\large\bf NEW THEORY OF SUPERFLUIDITY. } \vskip 1mm
\centerline {\large\bf METHOD OF EQUILIBRIUM DENSITY MATRIX } \vskip 7mm

\centerline {\large\bf Boris V. BONDAREV } \vskip 7mm

\centerline{\it Moscow Aviation Institute, Volokolamskoye Shosse 4,
125871, Moscow, Russia }\vskip 7mm

\centerline{E\hh -mail: bondarev.b@mail.ru }\vskip 7mm

\par\hskip 5mm{\it The variational theory of equilibrium boson system state to have been previously developed by the author under the density matrix formalism is applicable for researching equilibrium states and thermodynamic properties of the quantum Bose gas which consists of zero-spin particles. Particle pulse distribution function is obtained and duly employed for calculation of chemical potential, internal energy and gas capacity temperature dependences. It is found that specific phase transition, which is similar to transition of liquid helium to its superfluid state, occurs at the temperature exceeding that of the Bose condensation. } \vskip 7mm

\centerline {\bf 1. Introduction } \vskip 2mm

\par Today, the matter of Bose condensation aroused quite a lively interest in research of nature of this phenomenon and thermodynamic multifrequency system properties it occurs in [1-7].

\par The most general statistical formulation of the multifrequency system in quantum mechanics is gained by applying a density matrix. Specific variational method of tracing a boson equilibrium system density matrix being sought in the principle of minimum free energy is suggested in [8, 9]. The method under this paper is applied for formulation of thermodynamic properties of quantum Bose gas.

\vskip 4mm
\centerline {\bf 2.	Uniform Distribution of Particles in Space } \vskip 2mm

\par For analysis of equilibrium states of quantum gas consisting of angular momentum zero-spin particles we apply a variational density matrix method. We assume that $N$ of the above particles finds itself within a certain space area with the volume to be equal to $V$. If the particles have their mean uniform distribution, a single-particle density matrix, which in coordinate representation depends on radius vectors $\bf r$ and
${\bf r}^\pr$, will obtain the form as follows:
$$	\varrho_{\h{\bf r}\h{\bf r}^\pr}=\frac{1}{V}\hh\sum\limits_{\bf k}\hh p_{\h\bf k}\hs\exp\h\Bigl(\hh i\hs{\bf k}\hs({\bf r}-{\bf r}^\pr)
\Bigr)\hh , \eqno (2.1) $$
where $p_{\h\bf k}$ is the probability that an arbitrarily accepted particle is in the state defined by wave vector $\bf k$. Particle momentum -- ${\bf p}=\hbar\hh{\bf k}$. Probability $p_{\h\bf k}$ meets its normalizing condition,
$$ \sum\limits_{\bf k}\hh p_{\h\bf k}=1\hh . \eqno (2.2) $$

\par As formula (2.1) indicates, the momentum representation density matrix is diagonal, i.e.
$$ \varrho_{\h{\bf k}\h{\bf k}^\pr}=p_{\h\bf k}\hs
\delta_{\h{\bf k}\h{\bf k}^\pr}\hh , $$
and transition from momentum representation to the nodal one is effected by an unitary matrix
$$ u_{\h{\bf r}\h{\bf k}}=\frac{1}{\sqrt{V}}\hs
\exp\h(\hh i\hh{\bf k}\hh{\bf r})\hh , \eqno (2.3) $$
which is subject to the condition as follows:
$$ \int\limits_V u_{\h{\bf r}\h{\bf k}}\hs u_{\h{\bf r}\h{\bf k}}^*\hs
{\rm d}{\bf r}=\delta_{\h{\bf k}\h{\bf k}^\pr} \hskip 15mm
\hbox{or} \hskip 15mm
\frac{1}{V}\int\limits_V e^{\h i\h({\bf k}-{\bf k}^\pr)\h{\bf r}}\hs
{\rm d}{\bf r}=\delta_{\h{\bf k}\h{\bf k}^\pr}\hh . $$

\vskip 4mm
\centerline {\bf 3.	Kinetic Energy of Particles } \vskip 2mm

\par Particle kinetic energy operator $\hat H^{(1)}$ takes the form as follows:
$$ \hat H^{(1)}=\hh -\hs\frac{\hbar^{\h 2}\hh\nabla^{\h 2}}{2\hh m}\hh , \eqno (3.1) $$
where $m$ – particle mass.

\par Applying formula:
$$ H_{\h{\bf k}\h{\bf k}^\pr}=\int\limits_V u_{\h{\bf r}\h{\bf k}}^*\hs
\hat H^{(1)}\hs u_{\h{\bf r}\h{\bf k}^\pr}\hs{\rm d}{\bf r} $$
we can find elements of operator $\hat H^{(1)}$ in momentum representation:
$$ H_{\h{\bf k}\h{\bf k}^\pr}=\varepsilon_{\h\bf k}\hs
\delta_{\h{\bf k}\h{\bf k}^\pr}\hh . \eqno (3.2) $$
where $\varepsilon_{\h\bf k}$ is the particle kinetic energy with pulse
${\bf p}=\hbar\hh{\bf k}$:
$$ \varepsilon_{\h\bf k}=\frac{\hbar^{\h 2}\hh k^{\h 2}}{2\hh m}\hh .
\eqno (3.3) $$

\makeatletter\renewcommand{\@evenhead} {\vbox{\hbox
to\textwidth{\rm\thepage\hfil\it New theory of superfluidity. Method of equilibrium density  matrix \hfil}\vskip 1mm\hrule
\vskip -3mm}}

\makeatletter\renewcommand{\@oddhead} {\vbox{\hbox
to\textwidth{\hfill\it Bondarev B.V. \hfill\rm\thepage}\vskip 1mm\hrule\vskip -3mm}}

\vskip 4mm
\centerline {\bf 4. Particle Interaction Energy } \vskip 2mm
	
\par We assume that $U_{{\bf r}_1{\bf r}_2}=U_{{\bf r}_1-{\bf r}_2}$ is potential two particle interaction energy. In virtue of translation invariance the potential energy depends on vectors difference ${\bf r}_1-{\bf r}_2$ only and it is represented by its symmetrical function:
$$ U_{{\bf r}_1{\bf r}_2}=U_{{\bf r}_2{\bf r}_1}\hh . $$
In this case, interaction Hamiltonian matrix elements shall be formulated by the following method:
$$ H_{{\bf r}_1{\bf r}_2,\h{\bf r}_1^\pr{\bf r}_2^\pr}=
\frac{1}{2}\hs U_{{\bf r}_1{\bf r}_2}\hs(
\delta_{\h{\bf r}_1\h{\bf r}_1^\pr}\hs
\delta_{\h{\bf r}_2\h{\bf r}_2^\pr}+
\delta_{\h{\bf r}_1\h{\bf r}_2^\pr}\hs
\delta_{\h{\bf r}_2\h{\bf r}_1^\pr})\hh , \eqno (4.1) $$
where $\delta_{\h{\bf r}_1\h{\bf r}_1^\pr}$ is a delta-Dirac function.

\par We can find interaction Hamiltonian matrix elements in momentum representation using the following expression:
$$ H_{{\bf k}_1{\bf k}_2,\h{\bf k}_1^\pr{\bf k}_2^\pr}=
\int\limits_V u_{\h{\bf r}_1\h{\bf k}_1}^*\hs
u_{\h{\bf r}_2\h{\bf k}_2}^*\hs
H_{{\bf r}_1{\bf r}_2,\h{\bf r}_1^\pr{\bf r}_2^\pr}\hs
u_{\h{\bf r}_1^\pr\h{\bf k}_1^\pr}\hs
u_{\h{\bf r}_2^\pr\h{\bf k}_2^\pr}\hs
{\rm d}{\bf r}_1\hs{\rm d}{\bf r}_2\hs
{\rm d}{\bf r}_1^\pr\hs{\rm d}{\bf r}_2^\pr\hh . $$
Substitution of matrices (2.3) and expression (4.1) into the formula makes it possible to obtain the expression as follows:
$$ H_{{\bf k}_1{\bf k}_2,\h{\bf k}_1^\pr{\bf k}_2^\pr}=
\frac{1}{2\hh V}\hs\delta({\bf k}_1+{\bf k}_2-{\bf k}_1^\pr -
{\bf k}_2^\pr)\hs\int\limits_V U_{\bf r}\hs\Bigl(
e^{\h i\h({\bf k}_1-{\bf k}_1^\pr)\h{\bf r}}+
e^{\h i\h({\bf k}_2-{\bf k}_2^\pr)\h{\bf r}}
\Bigr)\hs{\rm d}{\bf r}\hh . \eqno (4.2) $$
In this case, two particle and wave ${\bf k}_1$ и ${\bf k}_1$ vector interaction energy will be represented by function [8]
$$ V_{{\bf k}_1{\bf k}_2}=(2-\delta_{\h{\bf k}_1{\bf k}_2})\hs
H_{{\bf k}_1{\bf k}_2,\h{\bf k}_1{\bf k}_2}=
\frac{1}{2\hh V}\hs(2-\delta_{\h{\bf k}_1{\bf k}_2})\hh
\int\limits_V U_{\bf r}\hs\Bigl(1+e^{\h i\h({\bf k}_1-{\bf k}_2)\h{\bf r}}
\Bigr)\hs{\rm d}{\bf r}\hh . \eqno (4.3) $$
This kind of an expression can be represented by the following form:
$$ V_{{\bf k}_1{\bf k}_2}=
\frac{1}{ V}\hs\bigl[\hh(2-\delta_{\h{\bf k}_1{\bf k}_2})\hh U+
v_{\h{\bf k}_1-{\bf k}_2}\bigr]\hh , \eqno (4.4) $$
where
$$ U=\int\limits_V U_{\bf r}\hs{\rm d}{\bf r}\hh , \eqno (4.5) $$
$$ v_{\h{\bf k}_1-{\bf k}_2}=-\int\limits_V U_{\bf r}\hs\Bigl(
1-e^{\h i\h({\bf k}_1-{\bf k}_2)\h{\bf r}}\Bigr)\hs{\rm d}{\bf r}\hh .
\eqno (4.6) $$

\par Any real closely interspaced particles tend to be exposed to strong repulsion. Therefore, it can be assumed that all the values of function (4.6) are rather small against the value of function (4.5) which will be considered as a positive one. It is believed that gas properties could be kept out of their significant change, provided that particle interaction is represented by the following simplified formula:
$$ V_{{\bf k}_1{\bf k}_2}=
\frac{1}{ V}\hs(2-\delta_{\h{\bf k}_1{\bf k}_2})\hh U\hh ,
\eqno (4.7) $$
where $U>0$.

\vskip 4mm
\centerline {\bf 5. Gas Internal Energy } \vskip 2mm

\par As stated in the approximation of statistically independent particles, internal energy is defined by expression [8]:
$$ E=\sum\limits_{\bf k}\hh\varepsilon_{\bf k}\hs\bar n_{\bf k}+
\frac{1}{2}\hh\sum\limits_{{\bf k}_1}\hh\sum\limits_{{\bf k}_2}\hh
V_{{\bf k}_1{\bf k}_2}\hs\bar n_{{\bf k}_1}\hs\bar n_{{\bf k}_2}\hh ,
\eqno (5.1) $$
where $\bar n_{{\bf k}_1}=N\hh p_{\h\bf k}$ is a mean number of particles with wave vector $\bf k$,
$$ \sum\limits_{\bf k}\hh\bar n_{\bf k}=N\hh . \eqno (5.2) $$

\par Substitution of expression (4.4) into formula (5.1) makes it possible to modify it into:
$$ E=\sum\limits_{\bf k}\hh\varepsilon_{\bf k}\hs\bar n_{\bf k}+
\frac{1}{2\hh V}\hh\biggl(2\hh U\hh N^{\h 2}-
U\hh\sum\limits_{\bf k}\hh\bar n^{\h 2}_{\bf k}+
\sum\limits_{{\bf k}_1}\hh\sum\limits_{{\bf k}_2}\hh
v_{\h{\bf k}_1-{\bf k}_2}\hs\bar n_{{\bf k}_1}\hs\bar n_{{\bf k}_2}
\biggr)\hh . \eqno (5.3) $$

\par If $v_{\h{\bf k}_1-{\bf k}_2}=0$, the above form fits to the Huang-Yang-Luttinger model [1].

\vskip 4mm
\centerline {\bf 6. Particle Pulse Distribution Function } \vskip 2mm

\par As stated in the approximation of statistically independent particles, the equation of function $\bar n_{\bf k}$ which describes particle pulse distribution process takes up the following form [8]:
$$ \vartheta\hs\ln\hh\frac{1+\bar n_{\bf k}}
{\bar n_{\bf k}}=\bar\varepsilon_{\bf k}-\mu\hh , \eqno (6.1) $$
where $\vartheta=k_{\hh\rm B}\hh T$, average particle energy is formed as:
$$ \bar\varepsilon_{\h\bf k}=
\frac{\partial E}{\partial\h\bar n_{\bf k}}=\varepsilon_{\h\bf k}+\frac{1}{V}\hh
\biggl(-\hs U\hs\bar n_{\bf k}+\sum\limits_{{\bf k}^\pr}\hh
v_{\h{\bf k}-{\bf k}^\pr}\hs\bar n_{{\bf k}^\pr}
\biggr)\hh . \eqno (6.2) $$

\par We regard values $v_{\h{\bf k}-{\bf k}^\pr}$, as small perturbations:
$v_{\h{\bf k}-{\bf k}^\pr}\ll U$ and we can find the distribution function using the formula as follows:
$$ \bar n_{\bf k}=f\bigl[\hh\beta\hh(\varepsilon_{\bf k}-\mu)\hh
\bigr]+\delta\bar n_{\bf k}\hh , \eqno (6.3) $$
where $\beta=1/\vartheta$, $f(\xi)$ -- one-variable function
$$ \xi=\beta\hh(\varepsilon_{\bf k}-\mu)\hh . \eqno (6.4) $$
Upon some simple transformations, substitution of expression (6.3) into equation (6.1) makes it possible to equate function $f(\xi)$:
$$ \ln\h\frac{1+f}{f}=\xi-b\hs f\hh , \eqno (6.5) $$
where
$$ b=\frac{\beta\hs U}{V}\hh , $$
and to produce the following formula:
$$ \delta\bar n_{\bf k}=-\hs\frac
{\beta\hh(1+f_{\bf k})\hh f_{\bf k}}
{V-\beta\hs U\hh(1+f_{\bf k})\hh f_{\bf k}}\hs
\sum\limits_{{\bf k}^\pr}\hh v_{\h{\bf k}-{\bf k}^\pr}\hs
f_{{\bf k}^\pr}\hh , \eqno (6.6) $$
where
$$ f_{\bf k}=f\bigl[\hh\beta\hh(\varepsilon_{\bf k}-\mu)\hh\bigr]\hh . $$
For graph of function $f=f(\xi)$ to be the solution of equation (6.5), as applied to one of the values of parameter $b$, see Fig. 1. The curve has two asymptotes: horizontal one: $f=0$ and oblique one: $f=\xi/b$. Function $f=f(\xi)$ can be suitable for the positive values of argument $\xi$ only, like:
$$ \xi\geq \xi_{\rm o}>0\hh , $$
where the least value of $\xi_{\rm o}$ depends on parameter $b$. Value $b$ within the thermodynamic limit $(V\hh\to\hh\infty)$ becomes zero at any temperature range. In this case,
$$ \lim\limits_{b\hh\to\hh 0}\hh \xi_{\rm o}=0\hh . $$

\unitlength=1.4mm \centerline{\begin{picture}(51,40)
\put(0,3){0}\put(8,3){$\xi_o$}
\put(1,7){\vector(1,0){47.5}}\put(47.5,3){$\xi$}
\put(1,7){\vector(0,1){29}}\put(2.7,34){$f(\xi)$} \put(1,7.1){
\multiput(8.40,0)(0,1.7){4}{\line(0,1){1.3}} \special{em:linewidth
0.3pt} \put(48.06,0.08) {\special{em:moveto}} \put(36.04,0.16)
{\special{em:moveto}} \put(29.62,0.24) {\special{em:lineto}}
\put(25.46,0.32) {\special{em:lineto}} \put(22.48,0.40)
{\special{em:lineto}} \put(20.22,0.48) {\special{em:lineto}}
\put(18.45,0.56) {\special{em:lineto}} \put(17.02,0.64)
{\special{em:lineto}} \put(15.84,0.72) {\special{em:lineto}}
\put(14.86,0.80) {\special{em:lineto}} \put(12.98,1.00)
{\special{em:lineto}} \put(11.40,1.26)   {\special{em:lineto}}
\put(10.15,1.58)   {\special{em:lineto}} \put(9.25,1.99)
{\special{em:lineto}} \put(8.684,2.49)   {\special{em:lineto}}
\put(8.468,3.13)   {\special{em:lineto}} \put(8.620,3.93)
{\special{em:lineto}} \put(9.164,4.93)   {\special{em:lineto}}
\put(10.16,6.19)   {\special{em:lineto}} \put(11.91,8.00)
{\special{em:lineto}} \put(16.30,12.0) {\special{em:lineto}}
\put(20.98,16.0) {\special{em:lineto}} \put(25.78,20.0)
{\special{em:lineto}} \put(30.64,24.0) {\special{em:lineto}}
\put(35.58,28.0) {\special{em:lineto}} } \end{picture}}

\vskip 1mm
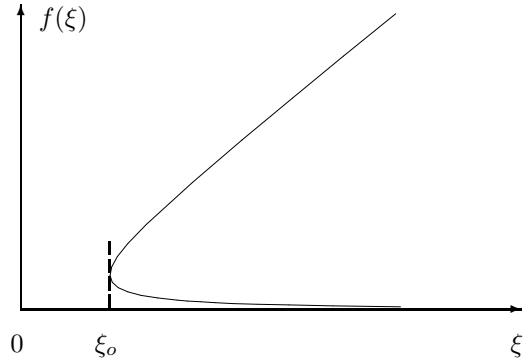
\centerline{\it Fig. 1. Distribution function $f=f(\xi)$, as applied for solution of equation \rm (6.5) } \vskip 3mm

\par At any value of $\xi_{\rm o}>0$, function $f=f(\xi)$ takes up two values with the minor one defined as $f_n(\xi)$ and with the greater one -- $f_c(\xi)$. As provided by equation (6.5), if specific macroscopic values of volume $V$ are applied, these values are defined by the following expressions:
$$ f_n(\xi)=\frac{1}{e^{\hh\xi}-1}\hh , \eqno (6.7) $$
$$ f_c(\xi)=\frac{V\hs\xi}{\beta\hs U}\equiv
\frac{N\hs\xi}{\beta\hs c\hs U}\hh , \eqno (6.8) $$
where $c=N/V$ – particle concentration.

\par It should be noted that an average number of particles $f_c(\xi)$ is macroscopically great against number $f_n(\xi)$.

\vskip 4mm
\centerline {\bf 7.	Chemical Potential } \vskip 2mm

\par According to formula (6.3) the particle pulse distribution stated in its first approximation is described by the following function:
$$ \bar n_{\bf k}=f\biggl(\beta\hs(\varepsilon_{\bf k}-\mu),\hh
\frac{N}{\beta\hs c\hs U}\biggr)\h , \eqno (7.1) $$
which parametrically contains temperature, chemical potential and dimensionless quantity $N/\beta\hh c\hh U$. Substitution of function (7.1) into equation (5.2) results in the following equation:
$$ \sum\limits_{\bf k}\hh f\biggl(\beta\hs(\varepsilon_{\bf k}-\mu),\hh
\frac{N}{\beta\hs c\hs U}\biggr)=N\hh , \eqno (7.2) $$
which implicitly sets chemical potential dependence on temperature.

\par Should certain function $\varphi_{\hh\bf k}$ of argument $\bf k$ be of continuous nature, the summation by wave vectors $\bf k$ may be substituted for the following integration:
$$ \sum\limits_{\bf k}\hh\varphi_{\hh\bf k}=
\frac{V}{(2\hh\pi)^{\h 3}}\hs\int\varphi_{\hh\bf k}\hs{\rm d}{\bf k}\hh . $$
If function $\varphi_{\hh\bf k}$ is a complex one for vector $\bf k$, where kinetic energy $\varepsilon_{\bf k}$ acts as an intervening variable, the foregoing equation integral may be transformed with formula (3.3) by the method as follows:
$$ \int\varphi(\varepsilon_{\h\bf k})\hs{\rm d}{\bf k}=
\frac{4\hh\pi\hh m\hh\sqrt{2\hh m}}{\hbar^{\h 3}}\hs
\int\limits_0^\infty\phi(\varepsilon)\hs\sqrt\varepsilon\hs
{\bf d}\varepsilon\hh . $$

\par As s result, we obtain the known formula:
$$ \sum\limits_{\bf k}\hh\varphi(\varepsilon_{\h\bf k})=A\hs N\hh
\int\limits_0^\infty\phi(\varepsilon)\hs\sqrt\varepsilon\hs
{\bf d}\varepsilon\hh , \eqno (7.3) $$
where
$$ A=\frac{m\hh\sqrt{2\hh m}\hs V}
{2\hh\pi^{\h 2}\hh\hbar^{\h 3}\hh N}\hh . \eqno (7.4) $$

\par We assume that at high temperatures distribution of particles by their states is described by dependence (6.7). In this case, condition (7.2) may be formulated (7.3) by the method as follows:
$$ A\hh\int\limits_0^\infty\frac
{\sqrt\varepsilon\hs{\rm d}\varepsilon}
{e^{\h\beta\h(\varepsilon-\mu)}-1}=1\hh . $$
If a new integration variable is applied, in particular:
$$ y=\beta\hh\varepsilon $$
this equation may be transformed into the following form:
$$ A\hs\vartheta^{\h 3/2}\hs J\biggl(-\hs\frac{\mu}{\vartheta}\biggr)=1
\hh , \eqno (7.5) $$
where
$$ J(x)=\int\limits_0^\infty\frac{\sqrt y\hs{\rm d}y}
{e^{\h x+y}-1}\hh . \eqno (7.6) $$

\par As it is known, temperature value  , when chemical potential becomes zero, is called Bose condensation temperature:
$$ \vartheta_B=\frac{1}{\bigl(A\hs J(0)\bigr)^{2/3}}=
\frac{\hbar^{\h 2}}{m}\biggl(\frac{\sqrt 2\hs\pi^{\h 2}\hh N}
{V\hs J(0)}\biggr)^{2/3}\hh . \eqno (7.7) $$

\par We apply dimensionless quantities:
$$ \tau=\frac{\vartheta}{\vartheta_B}\hh , \hskip 15mm \tilde\mu=\frac{\mu}{\vartheta_B}\hh . \eqno (7.8) $$

\par Now, equation (7.5) may be represented in the form suitable for numerical solution:
$$ \tau^{\h 3/2}\hs J\biggl(-\hs\frac{\tilde\mu}{\tau}\biggr)=J(0)
\hh , \eqno (7.9) $$
This kind of equation defines dependence $\tilde\mu=\tilde\mu(\tau)$ which can be significant when $\tau\geq 1$ only.

\par If $\vartheta<\vartheta_B$, equation (7.5) is not solvable and should be substituted for another one. We can obtain the above new equation by the method as follows.

\par We assume that at temperatures below specific critical value $\vartheta_c$ the distribution function takes up the following form:
$$ \bar n_{\bf k}=\left\{\begin{array}{ccc}
f_n\bigl[\hh\beta\hh(\varepsilon_{\bf k}-\mu)\hh\bigr]
& \hbox{at} & {\bf k}\neq 0 \hs , \medskip \\
f_c(-\hh\beta\hh\mu) & \hbox{at} & {\bf k}=0\hs . \\
\end{array}\right. \eqno (7.10) $$

\par With this function substituted in normalizing condition (7.2), the following equation can be obtained:
$$ N_n+N_c=N\hh , \eqno (7.11) $$
where value
$$ N_n=\sum\limits_{{\bf k}\hh\neq\hh 0}
f_n\bigl[\hh\beta\hh(\varepsilon_{\bf k}-\mu)\hh\bigr] \eqno (7.12) $$
will be named as a normal state particle number and value
$$ N_c=f_c(-\hh\beta\hh\mu) \eqno (7.13) $$
-- as a condensed state particle number. The above particles produce zero pulse and so-called condensate. On substituting expressions (6.7) and (6.8) in formulas (7.12) and (7.13) and changing sum $\bf k$ for energy $\varepsilon$ integral in (7.12), we transform equation (7.11) into the following form:
$$ \frac{\tau^{\h 3/2}}{J(0)}\hs J\biggl(-\hs\frac{\tilde\mu}{\tau}\biggr)-\frac{\tilde\mu}{I}=1\hh , \eqno (7.14) $$
where $I$ -- particle interaction parameter,
$$ I=\frac{c\hs U}{\vartheta_B}\hh . $$

\par For graphs of functions $\tilde\mu=\tilde\mu(\tau)$ to be the solutions of equations (7.9) and (7.14), as applied to various values of parameter $I$, see Fig. 2. If $I=0$, we produce the known chemical potential temperature dependence applying the theory of boson interaction Bose condensation. If $I\neq 0$, the nature of chemical potential dependence on temperature is significantly changed. If $\tau=0$, chemical potential $\tilde\mu(0)$ is equal to $-\hs I$. Function $\tilde\mu(\tau)$ is monotonically increased within the range of zero to critical value $\tau>1$. If $\tau\in(1,\hh\tau_c)$, function $\tilde\mu(\tau)$ starts zigzagging accepting three values from the interval against every value $\tau$ and joining to function $\tilde\mu(\tau)$ of respective $I=0$ at point $(1,\hh 0)$. Value $\tau_c$ is a monotonically increased function of parameter $I$. Should the value concerned get duly defined, critical temperature $\vartheta_c=\tau_c\hh\vartheta_B$ will be produced.

\vskip 4mm
\centerline {\bf 8. Order Parameter } \vskip 2mm

\par If ${\bf k}=0$, particle macroscopic number transition in its condensed state shall be referred to specific change of phase. As provided by the phase transition theories, the level of particle phase transformation is conventionally characterized by its $\eta$-order parameter. In the case analyzed, the order parameter shall be defined, as the ratio of condensed state particle number $N_c$ to the full number of particles $N$:
$$ \eta=\frac{N_c}{N}\hh . \eqno (8.1) $$

\par When formulae (6.12) and (7.13) are applied, it is not difficult to find the relationship between chemical potential and order parameter:
$$ \mu=-\hs c\hs U\hh\eta\hskip 10mm \hbox{or} \hskip 10mm
\tilde\mu=-\hs I\hh\eta \hskip 10mm \hbox{at} \hskip 10mm
\vartheta\leq\vartheta_c\hh . \eqno (8.2) $$

\par Substitution of second expression (8.2) into formula (7.4) makes it possible to obtain the equation as follows:
$$ \frac{\tau^{\h 3/2}}{J(0)}\hs J\biggl(\frac{I\hs\eta}{\tau}\biggr)+\eta=1
\hh , \eqno (8.3) $$

\par For graphs of function $\eta=\eta(\tau)$ to be the solution of this equation, see Fig. 3.

\par If $I=0$, equation (8.3) produces the dependence as follows:
$$ \eta=1-\tau^{\h 3/2}\hh , \eqno (8.4) $$
which describes particle interaction Bose condensation.

\unitlength=1mm \centerline{\begin{picture}(88,59)
\put(71,35){\it 1} \put(31,38){\it 2}\put(31,28){\it 3}\put(31,18){\it 4}
\put(31,8){\it 5}
\put(12,45){\vector(1,0){74.5}}\put(84.5,47.5){$\tau$}
\put(32,45){\line(0,1){1}}\put(29.7,47.5){0.5}
\put(52,45){\line(0,1){1}}\put(49.7,47.5){1.0}
\put(72,45){\line(0,1){1}}\put(69.7,47.5){1.5}
\put(12,3){\vector(0,1){52}}\put(3,53){$\tilde\mu\h (\tau)$}
\put(12,45){\line(-1,0){1}}\put(3,44){0}
\put(12,35){\line(-1,0){1}}\put(0,34){$-\hh 0.25$}
\put(12,25){\line(-1,0){1}}\put(0,24){$-\hh 0.50$}
\put(12,15){\line(-1,0){1}}\put(0,14){$-\hh 0.75$} \put(12,
5){\line(-1,0){1}}\put(0,4) {$-\hh 1.0$}
\put(12,45){\unitlength=1mm\special{em:linewidth 0.3pt}
\put(40,0.00)      {\special{em:moveto}} \put(40,-0.0003)
{\special{em:moveto}} \put(40,-0.004)    {\special{em:lineto}}
\put(40.12,-0.02)  {\special{em:lineto}} \put(40.32,-0.065)
{\special{em:lineto}} \put(40.72,-0.159) {\special{em:lineto}}
\put(41.36,-0.335) {\special{em:lineto}} \put(42.24,-0.634)
{\special{em:lineto}} \put(43.32,-1.109) {\special{em:lineto}}
\put(44.56,-1.828) {\special{em:lineto}} \put(45.84,-2.865)
{\special{em:lineto}} \put(47.12,-4.312)   {\special{em:lineto}}
\put(48.28,-6,256)   {\special{em:lineto}} \put(49.12,-8.768)
{\special{em:lineto}} \put(49.52,-11.89)   {\special{em:lineto}}
\put(49.20,-15.57)   {\special{em:lineto}} \put(48.00,-19.66)
{\special{em:lineto}} \put(45.80,-23.91)   {\special{em:lineto}}
\put(42.60,-27.95)   {\special{em:lineto}} \put(38.65,-31.48)
{\special{em:lineto}} \put(34.30,-34.30) {\special{em:lineto}}
\put(29.93,-36.40) {\special{em:lineto}} \put(25.82,-37.80)
{\special{em:lineto}} \put(22.14,-38.72) {\special{em:lineto}}
\put(18.94,-39.29) {\special{em:lineto}} \put(16.22,-39.60)
{\special{em:lineto}} \put(13.94,-39.80) {\special{em:lineto}}
\put(0.000,-40.00) {\special{em:lineto}} }
\put(12,45){\unitlength=1mm\special{em:linewidth 0.3pt}
\put(40,0.00)      {\special{em:moveto}} \put(40,-0.0003)
{\special{em:moveto}} \put(40,-0.004)    {\special{em:lineto}}
\put(40.12,-0.02)  {\special{em:lineto}} \put(40.32,-0.065)
{\special{em:lineto}} \put(40.68,-0.159) {\special{em:lineto}}
\put(41.28,-0.334) {\special{em:lineto}} \put(42.08,-0.632)
{\special{em:lineto}} \put(43.04,-1.102) {\special{em:lineto}}
\put(44.08,-1.808) {\special{em:lineto}} \put(45.08,-2.818)
{\special{em:lineto}} \put(45.96,-4.204)   {\special{em:lineto}}
\put(46.56,-6,036)   {\special{em:lineto}} \put(46.64,-8.324)
{\special{em:lineto}} \put(46.08,-11.06)   {\special{em:lineto}}
\put(44.68,-14.14)   {\special{em:lineto}} \put(42.36,-17.35)
{\special{em:lineto}} \put(39.17,-20.44)   {\special{em:lineto}}
\put(35.32,-23.17)   {\special{em:lineto}} \put(31.15,-25.37)
{\special{em:lineto}} \put(27.02,-27.02) {\special{em:lineto}}
\put(23.16,-28.16) {\special{em:lineto}} \put(19.74,-28.90)
{\special{em:lineto}} \put(16.79,-29.36) {\special{em:lineto}}
\put(14.30,-29.65) {\special{em:lineto}} \put(12.21,-29.81)
{\special{em:lineto}} \put(10.47,-29.90) {\special{em:lineto}}
\put(0.000,-30.00) {\special{em:lineto}} }
\put(12,45){\unitlength=1mm\special{em:linewidth 0.3pt}
\put(40,0.00)      {\special{em:moveto}} \put(40,-0.0003)
{\special{em:moveto}} \put(40,-0.004)    {\special{em:lineto}}
\put(40.08,-0.02)  {\special{em:lineto}} \put(40.28,-0.064)
{\special{em:lineto}} \put(40.60,-0.159) {\special{em:lineto}}
\put(41.16,-0.334) {\special{em:lineto}} \put(41.80,-0.628)
{\special{em:lineto}} \put(42.52,-1.088) {\special{em:lineto}}
\put(43.20,-1.772) {\special{em:lineto}} \put(43.68,-2.730)
{\special{em:lineto}} \put(43.80,-4.008)   {\special{em:lineto}}
\put(43.40,-5,624)   {\special{em:lineto}} \put(42.28,-7.548)
{\special{em:lineto}} \put(40.32,-9.680)   {\special{em:lineto}}
\put(37.52,-11.87)   {\special{em:lineto}} \put(34.02,-13.94)
{\special{em:lineto}} \put(30.10,-15.71)   {\special{em:lineto}}
\put(26.08,-17.11)   {\special{em:lineto}} \put(22.27,-18.14)
{\special{em:lineto}} \put(18.84,-18.84) {\special{em:lineto}}
\put(15.87,-19.30) {\special{em:lineto}} \put(13.38,-19.59)
{\special{em:lineto}} \put(11.30,-19.76) {\special{em:lineto}}
\put(9.580,-19.87) {\special{em:lineto}} \put(8.164,-19.93)
{\special{em:lineto}} \put(6.992,-19.97) {\special{em:lineto}}
\put(0.000,-20.00) {\special{em:lineto}} }
\put(12,45){\unitlength=1mm\special{em:linewidth 0.3pt}
\put(40,0.00)      {\special{em:moveto}} \put(40,-0.0003)
{\special{em:moveto}} \put(40,-0.004)    {\special{em:lineto}}
\put(40.08,-0.02)  {\special{em:lineto}} \put(40.20,-0.064)
{\special{em:lineto}} \put(40.40,-0.158) {\special{em:lineto}}
\put(40.68,-0.330) {\special{em:lineto}} \put(40.92,-0.614)
{\special{em:lineto}} \put(41.00,-1.050) {\special{em:lineto}}
\put(40.68,-1.668) {\special{em:lineto}} \put(39.80,-2.488)
{\special{em:lineto}} \put(38.18,-3.494)   {\special{em:lineto}}
\put(35.72,-4,628)   {\special{em:lineto}} \put(32.50,-5.800)
{\special{em:lineto}} \put(28.72,-6.896)   {\special{em:lineto}}
\put(24.72,-7.824)   {\special{em:lineto}} \put(20.86,-8.544)
{\special{em:lineto}} \put(17.36,-9.060)   {\special{em:lineto}}
\put(14.34,-9.412)   {\special{em:lineto}} \put(11.84,-9.640)
{\special{em:lineto}} \put(9.784,-9.784) {\special{em:lineto}}
\put(8.120,-9.872) {\special{em:lineto}} \put(6.780,-9.924)
{\special{em:lineto}} \put(5.692,-9.956) {\special{em:lineto}}
\put(4.812,-9.980) {\special{em:lineto}} \put(4.092,-9.988)
{\special{em:lineto}} \put(3.500,-9.996) {\special{em:lineto}}
\put(0.000,-10.00) {\special{em:lineto}} }
\put(12,45){\unitlength=1mm\special{em:linewidth 0.3pt}
\put(40,0.00)      {\special{em:moveto}} \put(40.84,-0.163)
{\special{em:moveto}} \put(41.60,-0.333) {\special{em:lineto}}
\put(42.24,-0.507) {\special{em:lineto}} \put(42.84,-0.686)
{\special{em:lineto}} \put(43.40,-0.868) {\special{em:lineto}}
\put(44.68,-1.340) {\special{em:lineto}} \put(45.84,-1.834)
{\special{em:lineto}} \put(46.92,-2.346) {\special{em:lineto}}
\put(47.92,-2.875) {\special{em:lineto}} \put(48.88,-3.421)
{\special{em:lineto}} \put(49.80,-3.984) {\special{em:lineto}}
\put(50.72,-4,564) {\special{em:lineto}} \put(51.56,-5.156)
{\special{em:lineto}} \put(53.68,-6.712) {\special{em:lineto}}
\put(55.72,-8.360) {\special{em:lineto}} \put(57.68,-10.10)
{\special{em:lineto}} \put(59.60,-11.92) {\special{em:lineto}}
\put(61.52,-13.84) {\special{em:lineto}} \put(63.40,-15.85)
{\special{em:lineto}} \put(65.28,-17.95) {\special{em:lineto}}
\put(67.16,-20.15) {\special{em:lineto}} \put(69.00,-22.42)
{\special{em:lineto}} \put(70.88,-24.81) {\special{em:lineto}}
\put(72.72,-27.27) {\special{em:lineto}} \put(74.64,-29.85)
{\special{em:lineto}} } \end{picture}}

\centerline{\it Fig. 2. Chemical potential $\tilde\mu$, as a $\tau$-temperature function}
\centerline{\it for various values of interaction parameter $I$: }
\centerline{\it 1 -- $I=0$; 2 -- $I=0.25$; 3 -- $I=0.5$; 4 -- $I=0.75$; 5 -- $I=1.0$. } \vskip 2mm

\unitlength=1mm \centerline{\begin{picture}(80,65)
\put(43,23){\it
1} \put(51,23){\it 2}\put(55,23){\it 3}\put(58.5,23){\it
4}\put(61.5,23){\it 5}
\put(12,10){\vector(1,0){64.5}}\put(74.5,5.5){$\tau$}
\put(12,10){\line(0,-1){1}}\put(11,5){0}
\put(32,10){\line(0,-1){1}}\put(29.7,5){0.5}
\put(52,10){\line(0,-1){1}}\put(49.7,5){1.0}
\put(12,10){\vector(0,1){50}}\put(3,58){$\eta\hh (\tau)$}
\put(12,10){\line(-1,0){1}}\put(3,9){0}
\put(12,20){\line(-1,0){1}}\put(3,19){$0.25$}
\put(12,30){\line(-1,0){1}}\put(3,29){$0.50$}
\put(12,40){\line(-1,0){1}}\put(3,39){$0.75$}
\put(12,50){\line(-1,0){1}}\put(3,49) {$1.0$}
\put(12,10){\unitlength=1mm\special{em:linewidth 0.3pt}
\put(40,0.00)      {\special{em:moveto}} \put(40,0.0003)
{\special{em:moveto}} \put(40,0.004)    {\special{em:lineto}}
\put(40.12,0.02)  {\special{em:lineto}} \put(40.32,0.065)
{\special{em:lineto}} \put(40.72,0.159) {\special{em:lineto}}
\put(41.36,0.335) {\special{em:lineto}} \put(42.24,0.634)
{\special{em:lineto}} \put(43.32,1.109) {\special{em:lineto}}
\put(44.56,1.828) {\special{em:lineto}} \put(45.84,2.865)
{\special{em:lineto}} \put(47.12,4.312) {\special{em:lineto}}
\put(48.28,6,256) {\special{em:lineto}} \put(49.12,8.768)
{\special{em:lineto}} \put(49.52,11.89) {\special{em:lineto}}
\put(49.20,15.57) {\special{em:lineto}} \put(48.00,19.66)
{\special{em:lineto}} \put(45.80,23.91) {\special{em:lineto}}
\put(42.60,27.95) {\special{em:lineto}} \put(38.65,31.48)
{\special{em:lineto}} \put(34.30,34.30) {\special{em:lineto}}
\put(29.93,36.40) {\special{em:lineto}} \put(25.82,37.80)
{\special{em:lineto}} \put(22.14,38.72) {\special{em:lineto}}
\put(18.94,39.29) {\special{em:lineto}} \put(16.22,39.60)
{\special{em:lineto}} \put(13.94,39.80) {\special{em:lineto}}
\put(11.00,39.90) {\special{em:lineto}} \put(9.000,39.95)
{\special{em:lineto}} \put(7.000,40.00) {\special{em:lineto}}
\put(0.000,40.00) {\special{em:lineto}} }
\put(12,10){\unitlength=1mm\special{em:linewidth 0.3pt}
\put(40,0.00)     {\special{em:moveto}} \put(40,0.000333)
{\special{em:lineto}} \put(40,0.005332) {\special{em:lineto}}
\put(40.12,0.271) {\special{em:lineto}} \put(40.32,0.086)
{\special{em:lineto}} \put(40.68,0.212) {\special{em:lineto}}
\put(41.28,0.446) {\special{em:lineto}} \put(42.08,0.842)
{\special{em:lineto}} \put(43.04,1.469) {\special{em:lineto}}
\put(44.08,2.410) {\special{em:lineto}} \put(45.08,3.757)
{\special{em:lineto}} \put(45.96,5.604) {\special{em:lineto}}
\put(46.56,8.048) {\special{em:lineto}} \put(46.64,11.10)
{\special{em:lineto}} \put(46.08,14.75) {\special{em:lineto}}
\put(44.68,18.85) {\special{em:lineto}} \put(42.36,23.14)
{\special{em:lineto}} \put(39.17,27.26) {\special{em:lineto}}
\put(35.32,30.90) {\special{em:lineto}} \put(31.15,33.83)
{\special{em:lineto}} \put(27.02,36.03) {\special{em:lineto}}
\put(23.16,37.54) {\special{em:lineto}} \put(19.74,38.53)
{\special{em:lineto}} \put(16.79,39.15) {\special{em:lineto}}
\put(14.30,39.53) {\special{em:lineto}} \put(12.21,39.74)
{\special{em:lineto}} \put(10.47,39.86) {\special{em:lineto}}
\put(0.000,40.00) {\special{em:lineto}} \put(0.000,40.00)
{\special{em:lineto}} }
\put(12,10){\unitlength=1mm\special{em:linewidth 0.3pt}
\put(40.00,0.0000) {\special{em:moveto}} \put(40.00,0.0005)
{\special{em:lineto}} \put(40.00,0.008) {\special{em:lineto}}
\put(40.08,0.041) {\special{em:lineto}} \put(40.28,0.129)
{\special{em:lineto}} \put(40.60,0.317) {\special{em:lineto}}
\put(41.16,0.667) {\special{em:lineto}} \put(41.80,1.255)
{\special{em:lineto}} \put(42.52,2.177) {\special{em:lineto}}
\put(43.20,3.543) {\special{em:lineto}} \put(43.68,5.460)
{\special{em:lineto}} \put(43.80,8.016) {\special{em:lineto}}
\put(43.40,11.25) {\special{em:lineto}} \put(42.28,15.10)
{\special{em:lineto}} \put(40.32,19.36) {\special{em:lineto}}
\put(37.52,23.74) {\special{em:lineto}} \put(34.02,27.87)
{\special{em:lineto}} \put(30.10,31.42) {\special{em:lineto}}
\put(26.08,34.22) {\special{em:lineto}} \put(22.27,36.27)
{\special{em:lineto}} \put(18.84,37.68) {\special{em:lineto}}
\put(15.87,38.60) {\special{em:lineto}} \put(13.38,39.18)
{\special{em:lineto}} \put(11.30,39.53) {\special{em:lineto}}
\put(9.580,39.74) {\special{em:lineto}} \put(8.164,39.86)
{\special{em:lineto}} \put(6.992,39.94) {\special{em:lineto}}
\put(5.000,40.00) {\special{em:lineto}} \put(0.000,40.00)
{\special{em:lineto}} }
\put(12,10){\unitlength=1mm\special{em:linewidth 0.3pt} \put(40,
0.000)  {\special{em:moveto}} \put(40,   0.001)
{\special{em:lineto}} \put(40.00,0.016)  {\special{em:lineto}}
\put(40.08,0.0812) {\special{em:lineto}} \put(40.20,0.2573)
{\special{em:lineto}} \put(40.40,0.6312) {\special{em:lineto}}
\put(40.68,1.318) {\special{em:lineto}} \put(40.92,2.458)
{\special{em:lineto}} \put(41.00,4.200) {\special{em:lineto}}
\put(40.68,6.672) {\special{em:lineto}} \put(39.80,9.952)
{\special{em:lineto}} \put(38.18,13.98) {\special{em:lineto}}
\put(35.72,18.51) {\special{em:lineto}} \put(32.50,23.20)
{\special{em:lineto}} \put(28.72,27.58) {\special{em:lineto}}
\put(24.72,31.30) {\special{em:lineto}} \put(20.86,34.18)
{\special{em:lineto}} \put(17.36,36.24) {\special{em:lineto}}
\put(14.34,37.65) {\special{em:lineto}} \put(11.84,38.56)
{\special{em:lineto}} \put(9.784,39.14) {\special{em:lineto}}
\put(8.120,39.49) {\special{em:lineto}} \put(6.780,39.70)
{\special{em:lineto}} \put(5.692,39.82) {\special{em:lineto}}
\put(4.812,39.92) {\special{em:lineto}} \put(4.092,39.95)
{\special{em:lineto}} \put(3.500,39.98) {\special{em:lineto}}
\put(0.00,40)     {\special{em:lineto}} }
\put(12,10){\unitlength=1mm\special{em:linewidth 0.3pt} \put(0,40)
{\special{em:lineto}} \put(2,39.55)  {\special{em:lineto}}
\put(4,38.74)  {\special{em:lineto}} \put(6,37.68)
{\special{em:lineto}} \put(8,36.42)  {\special{em:lineto}}
\put(10,35.00) {\special{em:lineto}} \put(12,33.43)
{\special{em:lineto}} \put(14,31.72) {\special{em:lineto}}
\put(16,29.88) {\special{em:lineto}} \put(18,27.92)
{\special{em:lineto}} \put(20,25.86) {\special{em:lineto}}
\put(22,23.69) {\special{em:lineto}} \put(24,21.41)
{\special{em:lineto}} \put(26,19.03) {\special{em:lineto}}
\put(28,16.57) {\special{em:lineto}} \put(30,14.02)
{\special{em:lineto}} \put(32,11.38) {\special{em:lineto}}
\put(34,8.652) {\special{em:lineto}} \put(36,5.852)
{\special{em:lineto}} \put(38,2.968) {\special{em:lineto}}
\put(40,0) {\special{em:lineto}} } \end{picture}}

\vskip -4mm

\centerline{\it Fig. 3. Order parameter $\eta$, as a $\tau$-temperature function }
\centerline{\it for various values of interaction parameter $I$: }
\centerline{\it 1 -- $I=0$; 2 -- $I=0.25$; 3 -- $I=0.5$;
4 -- $I=0.75$; 5 -- $I=1.0$. } \vskip 2mm

\vskip 4mm
\centerline {\bf 9. Gas Internal Energy Dependence on Temperature } \vskip 2mm

\par If we base on the particle pulse distribution function, we can find thermal dependence of gas internal energy by applying formula (5.3). As duly specified by the first approximation, i.e. less correction $\delta\bar n_{\bf k}$ and members containing $v_{\h{\bf k}-{\bf k}^\pr}$, we will apply the following form:
$$ E=\sum\limits_{\bf k}\hh\varepsilon_{\bf k}\hs\bar n_{\bf k}\hh -
\hh\frac{U}{2\hh V}\hh\sum\limits_{\bf k}\hh\bar n^{\h 2}_{\bf k}
\hh , \eqno (9.1) $$
where constant summand $U\hh N^{\h 2}/V$ is rejected for simplicity and function
$\bar n_{\bf k}$ is set by formula (7.10) subject to application for the purpose transformation of expression (9.1) into the form as follows:
$$ E=E_n+E_c\hh , \eqno (9.2) $$
where the first summand
$$ E_n=\sum\limits_{\bf k}\hh\varepsilon_{\h\bf k}\hs
f_n\bigl[\hh\beta\hh(\varepsilon_{\bf k}-\mu)\hh\bigr] \eqno (9.3) $$
is particle energy in normal state. If we substitute $\bf k$-summation for energy integration $\varepsilon$, the above summand may be described as follows:
$$ E_n=\frac{N\hh\vartheta_B}{J(0)}\hs\tau^{\h 5/2}\hs
K\biggl(\frac{I\hs\eta}{\tau}\biggr)\hh , \eqno (9.4) $$
where
$$ K(x)=\int\limits_0^\infty\hh\frac{y\hs\sqrt y\hs{\rm d}y}
{e^{\h x+y}-1}\hh . \eqno (9.5) $$

\par Formula (9.3) is applicable for the temperatures of below critical point, i.e. when  $\tau<\tau_c$. Should temperatures concerned exceed the one of Bose condensation, i.e. when $\tau>1$, formula (9.3) will be adequate to that as follows:
$$ E_n=\frac{N\hh\vartheta_B}{J(0)}\hs\tau^{\h 5/2}\hs
K\biggl(-\hs\frac{\tilde\mu}{\tau}\biggr)\hh , \eqno (9.6) $$
where function $\tilde\mu(\tau)$ is the solution of equation (7.9).

\par The second summand in formula (9.2)
$$ E_c=-\hs\frac{1}{2}\hs c\hs U\hs N\hs\eta^{\h 2} \eqno (9.7) $$
is condensed state particle energy.

\par Now, we apply dimensionless quantity
$$ \tilde\varepsilon=\frac{E}{N\hs\vartheta_B}\hh , \eqno (9.8) $$
which is rated as per one gas energy particle and measured in $\vartheta_B$. Dependence of specific energy $\tilde\varepsilon$ on temperature is formulated as follows:
$$ \tilde\varepsilon(\tau)=\frac{N\hh\vartheta_B}{J(0)}\hs
\tau^{\h 5/2}\hs K\biggl(\frac{I\hs\eta}{\tau}\biggr)-
\frac{1}{2}\hs I\hs\eta^{\h 2} \eqno (9.9) $$
at $\tau\leq\tau_c$.

\par If $I=0$, we produce the following formula:
$$ \tilde\varepsilon(\tau)=\frac{K(0)}{J(0)}\hs\tau^{\h 5/2}
\eqno (9.10) $$
at $\tau\leq 1$.

\par Besides, if $\tau\geq 1$, the following formula is produced from (9.6)
$$ \tilde\varepsilon_n(\tau)=\frac{\tau^{\h 5/2}}{J(0)}\hs
K\biggl(-\hs\frac{\tilde\mu}{\tau}\biggr) \eqno (9.11) $$
for particle specific energy in normal state.

\par For graph of function $\tilde\varepsilon=\tilde\varepsilon(\tau)$, as applied to interaction parameter value $I=1$, see Fig. 4. For $I=0$ dependence $\tilde\varepsilon=\tilde\varepsilon(\tau)$ is represented, as monotonically increasing function to be reduced to zero at $\tau=0$. If $\tau\in (1,\hh\tau_c)$, the graph of function $\tilde\varepsilon=\tilde\varepsilon(\tau)$ starts zigzagging at any
$I\neq 0$. Function $\tilde\varepsilon=\tilde\varepsilon(\tau)$ accepts three values for every value $\tau$ within this interval. If $\tau=1$, the graph of function
$\tilde\varepsilon=\tilde\varepsilon(\tau)$ is joined to that of similar function of respective $I=0$. The least energy state is a stable against other several possible macrosystem states. Therefore, real dependence $\tilde\varepsilon=\tilde\varepsilon(\tau)$ at point $\tau=\tau_c$ will take up specific discontinuity. This means that discontinuous decrease of gas internal energy shall occur at the temperature being dropped down to value $T_c$. As specified in the graph of order parameter dependence on temperature (see Fig. 3), macroscopic number of particles shall pass to condensed state at $T=T_c$. The rest normal-state particles shall condense, as far as the rate of temperature goes down. As a matter of fact, Bose gas condensation may be specified as heat release-accompanied phase transition. If , the specific heat of phase transition is defined by the rate of discontinuity of function $\tilde\varepsilon=\tilde\varepsilon(\tau)$ при $\tau=\tau_c$.

\unitlength=1mm \centerline{\begin{picture}(82,73)
\put(12,5){\vector(1,0){60.5}}\put(70.5,1){$\tau$}
\multiput(42,4.5)(0,2){19}{\line(0,1){1.4}}\put(41.2,1){1}
\multiput(49.4,4.5)(0,2){20}{\line(0,1){1.4}}\put(48.5,1){$\tau_c$}
\put(12,2){\vector(0,1){64}}\put(4,63){$\tilde\varepsilon\hh (\tau)$}
\put(12,25){\line(-1,0){1}}\put(7,24){$0$}
\put(12,15){\line(-1,0){1}}\put(3,14){$-\hh\frac{\hh I\hh}{2}$}
\put(12,5){\line(-1,0){1}}\put(4,4){$-\hh I$}
\multiput(11.3,25.5)(1.5,1.5){27}{
\unitlength=1mm\special{em:linewidth 0.3pt} \put(0,0)
{\special{em:moveto}} \put(1.1,1.1){\special{em:lineto}} } 
\put(12,25){\unitlength=1mm\special{em:linewidth 0.3pt}
\put(30.00,16.85)    {\special{em:moveto}} \put(32.10,19.24)
{\special{em:lineto}} \put(32.46,19.65)    {\special{em:lineto}}
\put(33.36,20.60)    {\special{em:lineto}} \put(33.84,20.92)
{\special{em:lineto}} \put(34.29,21.30)    {\special{em:lineto}}
\put(34.83,21.58)    {\special{em:lineto}} \put(35.01,21.72)
{\special{em:lineto}} \put(35.58,21.94)    {\special{em:lineto}}
\put(36.03,21.98)    {\special{em:lineto}} \put(36.12,21.92)
{\special{em:lineto}} \put(36.39,21.80)    {\special{em:lineto}}
\put(36.54,21.72)    {\special{em:lineto}} \put(36.66,21.48)
{\special{em:lineto}} \put(36.84,21.26)    {\special{em:lineto}}
\put(37.02,20.70)    {\special{em:lineto}} \put(37.11,19.93)
{\special{em:lineto}} \put(37.14,19.20)    {\special{em:lineto}}
\put(37.11,18.27)    {\special{em:lineto}} \put(37.02,17.38)
{\special{em:lineto}} \put(36.84,16.34)    {\special{em:lineto}}
\put(36.66,15.31)    {\special{em:lineto}} \put(36.39,14.27)
{\special{em:lineto}} \put(36.12,13.24)    {\special{em:lineto}}
\put(35.82,12.26)    {\special{em:lineto}} \put(35.46,11.18)
{\special{em:lineto}} \put(35.10,10.20)    {\special{em:lineto}}
\put(34.71,9.198)    {\special{em:lineto}} }
\put(12,25){\unitlength=1mm\special{em:linewidth 0.3pt}
\put(36.12,13.24)    {\special{em:moveto}} \put(35.22,10.55)
{\special{em:moveto}} \put(34.14,7.898)    {\special{em:lineto}}
\put(32.97,5.444)    {\special{em:lineto}} \put(31.74,3.318)
{\special{em:lineto}} \put(30.48,1.390)    {\special{em:lineto}}
\put(29.25,-0.274)    {\special{em:lineto}} \put(28.04,-1.706)
{\special{em:lineto}} \put(26.86,-2.930)    {\special{em:lineto}}
\put(25.73,-3.970)    {\special{em:lineto}} \put(22.67,-6.230)
{\special{em:lineto}} \put(20.09,-7.602)    {\special{em:lineto}}
\put(17.96,-8.442)    {\special{em:lineto}} \put(16.18,-8.964)
{\special{em:lineto}} \put(14.69,-9.304)    {\special{em:lineto}}
\put(13.44,-9.524)    {\special{em:lineto}} \put(12.37,-9.666)
{\special{em:lineto}} \put(11.45,-9.764)    {\special{em:lineto}}
\put(10.66,-9.836)    {\special{em:lineto}} \put(9.960,-9.878)
{\special{em:lineto}} \put(7.494,-9.976)    {\special{em:lineto}}
\put(5.997,-9.988)    {\special{em:lineto}} \put(5.001,-10.00)
{\special{em:lineto}} \put(4.287,-10.01)    {\special{em:lineto}}
\put(0.000,-10.01)    {\special{em:lineto}} }
\put(12,25){\unitlength=1mm\special{em:linewidth 0.3pt}
\put(30.0,16.85)  {\special{em:moveto}} \put(32.55,19.97)
{\special{em:lineto}} \put(34.38,22.22) {\special{em:lineto}}
\put(35.94,24.06) {\special{em:lineto}} \put(37.35,25.70)
{\special{em:lineto}} \put(38.67,27.16) {\special{em:lineto}}
\put(39.96,28.70) {\special{em:lineto}} \put(41.19,30.16)
{\special{em:lineto}} \put(43.56,32.84) {\special{em:lineto}}
\put(44.70,34.14) {\special{em:lineto}} \put(45.87,35.54)
{\special{em:lineto}} \put(46.98,36.62) {\special{em:lineto}}
\put(48.12,37.88) {\special{em:lineto}} \put(49.23,39.16)
{\special{em:lineto}} \put(50.37,40.48) {\special{em:lineto}} }
\end{picture}}

\vskip 2mm
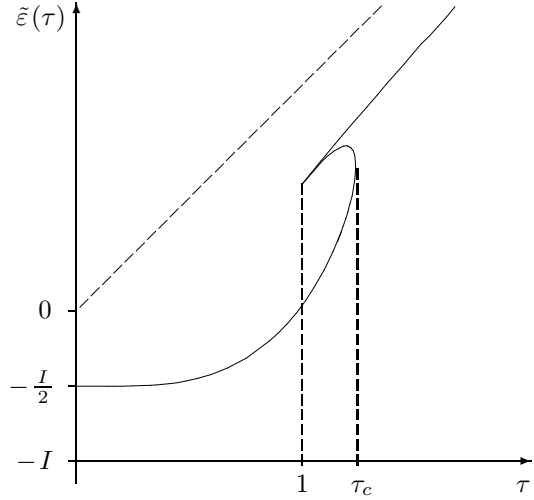
\centerline{\it Fig. 4. Bose gas internal energy $\tilde\varepsilon$, as $\tau$-temperature function }
\centerline{\it for interaction parameter value $I=1$ } \vskip 4pt

\vskip 4mm
\centerline {\bf 10. Heat Capacity of Gas } \vskip 2mm

\par As it is finally stated by the data above, we can use the following Bose gas specific energy formula:
$$ \tilde\varepsilon(\tau)=\left\{\begin{array}{ccl}
\ds\frac{\tau^{\h 5/2}}{J(0)}\hs K\biggl(\frac{I\hs\eta}{\tau}\biggr)-
\frac{1}{2}\hs I\hs\eta^{\h 2}
& \hbox{at} & \tau\leq\tau_c \hs , \medskip \\
\ds\frac{\tau^{\h 5/2}}{J(0)}\hs
K\biggl(-\hs\frac{\tilde\mu}{\tau}\biggr) & \hbox{at} & \tau\geq 1 \hs . \\ \end{array}\right. \eqno (10.1) $$
where dependence $\eta=\eta(\tau)$ is defined by equation (8.3) and dependence $\mu=\mu(\tau)$ -- by equation (7.9). Applying this formula we can find gas heat capacity dependence
$$ C_V=\frac{{\rm d}E}{{\rm d}T}$$
on temperature. As provided by ratio (9.8), i.e. $E=N\hh\vartheta_B\hh\tilde\varepsilon(\tau)$, the rate of heat capacity may be calculated by the following formula:
$$ C_V=N\hs k_B\hs\frac{{\rm d}\tilde\varepsilon}{{\rm d}\tau}\hh .
\eqno (10.2) $$

\par For graphs of dependence of derivative
$\frac{{\rm d}\tilde\varepsilon}{{\rm d}\tau}$ on temperature $\tau$, as applied to parameter $I=1$, see Fig. 5. If any value of $I>0$ is applied, the rate of Bose gas heat capacity for $\tau\in(0,\hh\tau_c)$ is monotonically increased from zero to infinity. If  $\tau>\tau_c$, particle interaction in the approximation involved does not affect temperature dependences of internal energy and gas heat capacity. Therefore, if $\tau>\tau_c$, the above dependences are similar to those produced by ideal Bose gas.

\vskip 4mm
\centerline {\bf 11. Energy Spectrum of Particles } \vskip 2mm

\par Now, let us find $\varepsilon_{\h\bf k}$-energy dependence of a particle on wave vector $\bf k$. Interaction of one particle with other gas particles makes energy get functions of the particle pulse distribution function. If stated in approximation of statistically independent particles, this functional is defined by formula (6.2). As provided by formulae (6.4) and (7.10), the equilibrium distribution function may be defined by the following expression:
$$ \bar n_{\h\bf k}=N_c\hs\delta_{\h\bf k}+
f_n\bigl[\hh\beta\hh(\varepsilon_{\h\bf k}-\mu)\hh\bigr]+
\delta\bar n_{\h\bf k}\hh . \eqno (11.1) $$

\par Substitution of this expression into formula (6.2) gives the following expression:
$$ \bar\varepsilon_{\h\bf k}=\varepsilon_{\h\bf k}+
c\hs\eta\hs(\hh -\hs U\hs\delta_{\h\bf k}+v_{\h\bf k})\hh .
\eqno (11.2) $$

\par While deriving this formula, it is taken into account that due to availability of a very small multiplier $1/V$ in formula (6.2) the macroscopic large bracketed summands only, i.e. those containing number $N_c$, may be specified as the substantial ones.

\unitlength=1mm \centerline{\begin{picture}(82,57)
\put(12,5){\vector(1,0){60.5}}\put(70.5,1.5){$\tau$}
\multiput(49.4,4.5)(0,2){23}{\line(0,1){1.4}}\put(48.5,1.5){$\tau_c$}
\put(12,5){\vector(0,1){45}}\put(13,45){$\ds\frac{C_V}{\hh\overline
N\hh k_B}$} \put(7,11.7){$\frac32$} \put(11,1.5){0} 
\put(12,5){\unitlength=1mm\special{em:linewidth 0.3pt}
\multiput(0,7.5)(2,0){30}{\line(1,0){1.4}} \put(36.90,44.94)
{\special{em:moveto}} \put(36.75,36.92)    {\special{em:lineto}}
\put(36.57,32.17)    {\special{em:lineto}} \put(36.36,28.80)
{\special{em:lineto}} \put(36.12,26.02)    {\special{em:lineto}}
\put(35.88,24.04)    {\special{em:lineto}} \put(35.61,22.19)
{\special{em:lineto}} \put(35.31,20.81)    {\special{em:lineto}}
\put(35.01,19.50)    {\special{em:lineto}} \put(34.71,18.48)
{\special{em:lineto}} \put(34.38,17.46)    {\special{em:lineto}}
\put(34.02,16.52)    {\special{em:lineto}} \put(34.71,18.48)
{\special{em:lineto}} \put(32.97,14.25)    {\special{em:lineto}}
\put(31.11,11.45)    {\special{em:lineto}} \put(29.25,9.395)
{\special{em:lineto}} \put(27.44,7.775)    {\special{em:lineto}}
\put(25.73,6.505)    {\special{em:lineto}} \put(24.14,5.480)
{\special{em:lineto}} \put(22.67,4.647)    {\special{em:lineto}}
\put(21.32,3.961)    {\special{em:lineto}} \put(20.09,3.394)
{\special{em:lineto}} \put(18.98,2.931)    {\special{em:lineto}}
\put(17.96,2.534)    {\special{em:lineto}} \put(17.33,2.307)
{\special{em:lineto}} \put(17.03,2.209)    {\special{em:lineto}}
\put(14.69,1.473)    {\special{em:lineto}} \put(12.71,0.9705)
{\special{em:lineto}} \put(11.18,0.6535)    {\special{em:lineto}}
\put(9.960,0.4529)    {\special{em:lineto}} \put(7.494,0.1597)
{\special{em:lineto}} \put(5.997,0.0518)    {\special{em:lineto}} }
\put(12,5){\unitlength=1mm\special{em:linewidth 0.3pt}
\put(37.30,8.909) {\special{em:moveto}} \put(38.67,8.701)
{\special{em:lineto}} \put(41.79,8.559) {\special{em:lineto}}
\put(44.70,8.462) {\special{em:lineto}} \put(47.55,8.383)
{\special{em:lineto}} \put(50.37,8.286) {\special{em:lineto}}
\put(53.16,8.224) {\special{em:lineto}} \put(55.98,8.166)
{\special{em:lineto}} \put(59.10,8.125) {\special{em:lineto}} }
\end{picture}}

\vskip 2mm
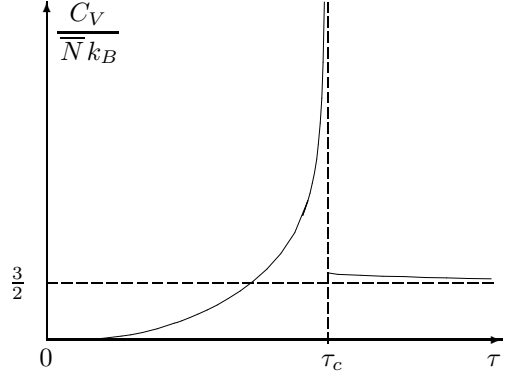
\centerline{\it Fig. 5. Bose gas heat capacity as a function of temperature $\tau$ }
\centerline{\it for interaction parameter value $I=1$. }

\vskip 4mm
\centerline {\bf 12. Superfluidity } \vskip 2mm

\par Now we consider Bose gas at the temperature of $T=0$ to be in the state described by a distribution function:
$$ \bar n_{\h\bf k}=N_1\hs\delta_{\h\bf kk_1}+N_2\hs\delta_{\h\bf kk_2}
\hh , \eqno (12.1) $$
where
$$ N_1+N_2=N\hh , \hskip 15mm {\bf k_1}\neq{\bf k_2}\hh . \eqno (12.2) $$

\par Formula (12.1) means that all the gas particles are distributed into two groups. Each particle of the first group has a pulse stated as ${\bf p_1}=\hbar\hh{\bf k_1}$. The particle number is equal to $N_1$. As for each particle in the second group, its pulse is stated as ${\bf p_2}=\hbar\hh{\bf k_2}$ and the number is equal to $N_2$.

\par Now, we find gas energy in this state by applying the following formula (9.1):
$$ E=\varepsilon_{\h\bf k_1}\hs N_1+\varepsilon_{\h\bf k_2}\hs N_2-
\frac{U}{2\hh V}\hh\bigl(N_1^2+N_2^2\bigr)\hh . \eqno (12.3) $$

\par We can prove that gas being in the state when all particles migrate at similar velocity remains stable, i.e. gas may be reconditioned under external effects only, but not spontaneously. This is possible when other gas states close to that under consideration have exceeded energy. Let us assume that number $N_1=N$ at a certain time $t_1$, i.e. all gas particles will have pulse $\bf p_1$. In this case, $N_2=0$, and gas energy is expressed by the following formula:
$$ E=\varepsilon_{\h\bf k_1}\hs N-
\frac{U}{2\hh V}\hs N\hh . \eqno (12.4) $$

\par Let us assume that at some time $t_2>t_1$ one or several particles $(N_2\geq 1)$ gain pulse $\bf p_2$ As a result, gas energy takes on value $E_2$ defined by formula (12.3), i.e. it increments.
$$ \triangle E=E_2-E_1=(\varepsilon_{\h\bf k_2}-\varepsilon_{\h\bf k_1})\hs N_2+
\frac{U}{V}\hs(N-N_2)\hs N_2\hh . \eqno (12.5) $$

\par If $N_2=N_{\rm o}$, function (12.5) has a maximum, where
$$ N_{\rm o}=\frac{1}{2}\hh\biggl(c+
\frac{\varepsilon_{\h\bf k_2}-\varepsilon_{\h\bf k_1}}{U}\biggr)\hh V
\hh . \eqno (12.6) $$

\par As $U>0$, function $\triangle E$ of $N_2$ will be positive for 
$N_2\in (1,\hh 2\hh N_{\rm o})$ provided that $N_{\rm o}>0$, i.e. when
$$ \varepsilon_{\h\bf k_1}<\varepsilon_{\h\bf k_2}+c\hs U\hh .
\eqno (12.7) $$

\par If $\bf k_2$, value $\triangle E$ takes on the least value like the function of 
${\bf k_2}=0$. Thus, to stop one of the particles within the flux it shall be duly energized
$$ \triangle E_1=-\hs\varepsilon_{\h\bf k_1}+\frac{N-1}{V}\hs U\hh . $$
This value will be positive, if
$$ \varepsilon_{\h\bf k_1}<\frac{N-1}{V}\hs U\hh , \eqno (12.8) $$
i.e. provided that ${\bf k_1<k}_c$, where the critical value is represented as follows:
$$ k_c=\frac{\sqrt{2\hs m\hs c\hs U}}{\hbar}\hh . $$
Respective critical value of speed will take the following form:
$$ v_c=\sqrt{\frac{2\hs c\hs U}{m}}\hh . $$
Condition (12.8) is stronger than condition (12.7), i.e. condition (12.7) is also executed for all $\bf k_1$ satisfying (12.8).

\par So, to take one or several $(1\leq N_2\leq 2\hh N_{\rm o})$ particles out of the flux with the particles running at velocity $v<v_c$, it is necessary to additionally energize them. Simultaneous and spontaneous transition of macroscopic number of particles from state $\bf k_1$ into particular state $\bf k_2$ is hardly probable. Since this kind of spontaneous transition is hardly probable, state of gas, where all particles migrate at similar velocity $v$, will be stable provided that $v<v_c$. This means that Bose gas exhibits its superfluidity property.

\par If required, the aforesaid arguments and calculations may be repeated, providing that gas temperature fits condition $0<T<T_c$. In this case, the least energy, when one of numbers $N_1$ or $N_2$ is equal to zero, provided that vectors ${\bf k}_1$ and 
${\bf k}_2\neq{\bf k}_1$ satisfy inequation (12.8), will also correspond to the state described by the following distribution function
$$ \bar n_{\h\bf k}=N_1\hs\delta_{\h\bf kk_1}+N_2\hs\delta_{\h\bf kk_2}+
f_n\bigl[\hh\beta\hh(\varepsilon_{\h\bf k}-\mu)\hh\bigr]\hh , $$
where $N_1+N_2=N$. Thus, we can summarize that critical velocity does not depend on temperature and the particles producing or having produced condensate tend to gain similar velocity. A “team” of particles prevents individual particles and large groups of particles from leaving off. So, if $v<v_c$, the condensate particles involved inside the directed movement and left to their own will migrate at their constant velocity as long as possible. But ganged particles exhibit joint directed motion to find out inside the space areas which dimensions do not exceed the coherence length. In this areas, particles move according to a quantum mechanics law. The particles spaced by a distance exceeding the coherence length may have different velocities distributed in space, as described by microscopic hydrodynamics laws.

\par Since directed motion of particles the superfluid gas component is formed of saves its stability against changes of individual particle velocities, any gas convection currents produced by temperature gradients will get free of attenuation over time. Such currents tend to cause condensate exhaustion. Zero-velocity particles may have very low fraction even at low temperatures produced at macroscopic space areas.

\vskip 4mm
\centerline {\bf 13. Conclusion } \vskip 2mm

\par Now, having written the last formula, the matter of theory applications is arisen. What kind of real multifrequency system could be described with the help of the obtained relations? It may be assumed that this system may be represented by liquid Не$^{\h 4}$. Likeness of liquid helium and quantum Bose gas thermal capacity temperature dependences argues in favor of this assumption. Herewith, the $\eta$-order parameter may be interpreted as relation of superfluid component densities $\varrho_c/\varrho$ to full liquid density. The temperature dependence of an order parameter, as specified in this paper, resembles experimental dependence of ratio $\varrho_c/\varrho$ on temperature. But the literature about liquid helium provides no information of discontinuity of internal energy being transitted into its superfluid state. It is known that medium field theories refer to the first approximation only accounting for the interacted particle effect produced on theromdynamic partlce properties. Perhaps, the internal energy jump at $T=T_c$ is caused by roughness of the accepted approximation and any correlations calculated would result in smoothing of gas internal energy temperature dependence.

\par This work is chiefly targeted to logical formulation of quantum Bose gas thermodynamic properties under the theory of density matix. The author suggests analyzing compliance of this model with any real multifrequency systems and matching theoretical relations with experimental results in an individual paper.

\vskip 4mm
\centerline {\bf Summary } \vskip 2mm

\par In summary, we hereby specify the following basic theory results:

\par 1.	The variational density matrix method recommended above is applied for formulation of quantum Bose gas equilibrium states and thermodynamic properties.

\par 2.	A single-frequency density matrix describing Bose gas equilibrium state is found in approximation of statistically independent particles.

\par 3.	Particle pulse distribution function is found. It is shown that this function exhibits specific features which can be interpreted as an effect caused by transition of Bose gas into its superfluid state at rather low temperatures.

\par 4.	Specific temperature dependences of chemical potential, order parameter, gas internal energy and its heat capacity are duly defined by applying the obtained particle pulse distribution function.

\vskip 5mm

\def\qhh{\hskip -5.7mm\qquad\parbox[t][1\height]{173mm}}

\centerline{\bf Refereces } \vskip 2mm

\noindent [1] \qhh{ K. Нuаng, C.N. Yang, J.M. Luttinger, {\it Imperfect Bose gas with hardspere unteractions}, Рhys. Rev., 1957, v. 105, p. 776-784. } \vskip 1.5mm

\noindent [2] \qhh{ K. Huang, Statistical mechanics, M.: Mir, 1966. }
\vskip 1.5mm

\noindent [3] \qhh{ D.R. Tilly, J. Tilly, Superfluidity and superconductivity, M.: Mir, 1977. } \vskip 1.5mm

\noindent [4] \qhh{ V.Z. Kresin, Superfluidity and superconductivity, M.: Science, 
1978. } \vskip 1.5mm

\noindent [5] \qhh{ M. Van den Berg, J.T. Lewis, J.V. Pule, {\it A general theory of Bose-Einstaein condensation}, Helv. Phys. Acta, 1986, v. 59, p. 1271-1288. }
\vskip 1.5mm

\noindent [6] \qhh{ M. Van den Berg, T.C. Dorlas, J.T. Kewis, J.V. Pule, {\it A perturbed mean gield model of an interacting boson gas and the large deviation pribciopke}, Commun. Math. Phys., 1990, v. 127, p. 41-69. } \vskip 1.5mm

\noindent [7] \qhh{ A. Minguzzu, S. Conti, M.P. Tosi, {\it The intеrnаl еnеrgy and соndеnsate fraction of a trapped interacting Bose gas}, J. Рhys.: Condens. Matter, 1997, v. 9, p. L33-L38. } \vskip 1.5mm

\noindent [8] \qhh{ B.V. Bondarev, {\it Density matrix method in theory of boson system equilibrium states}, Vestnik MAI, 1998, No. 1. } \vskip 1.5mm

\noindent [9] \qhh{ B.V. Bondarev, Density matrix method in quantum cooperative phenomena theory, M.: Sputnik+, 2001, p. 250. } \vskip 1.5mm

\end{document}